\documentclass[fleqn, usenatbib, letters]{mnras}

\usepackage{newtxtext,newtxmath}

\usepackage[T1]{fontenc}

\DeclareRobustCommand{\VAN}[3]{#2}
\let\VANthebibliography\thebibliography
\def\thebibliography{\DeclareRobustCommand{\VAN}[3]{##3}\VANthebibliography}

\usepackage{graphicx}	
\usepackage{amsmath}	

\title[A black hole imposter in NGC 1850]{NGC 1850 BH1 is another stripped-star binary masquerading as a black hole}

\author[El-Badry \& Burdge]{
Kareem El-Badry$^{1,2,3}$\thanks{E-mail: kareem.el-badry@cfa.harvard.edu}
and Kevin B. Burdge$^{4,5}$\\
$^{1}$Center for Astrophysics $|$ Harvard \& Smithsonian, 60 Garden Street, Cambridge, MA 02138, USA\\
$^{2}$Harvard Society of Fellows, 78 Mount Auburn Street, Cambridge, MA 02138\\
$^{3}$Max-Planck Institute for Astronomy, K\"onigstuhl 17, D-69117 Heidelberg, Germany\\
$^{4}$Department of Physics, Massachusetts Institute of Technology, Cambridge, MA 02139, USA\\
$^{5}$Kavli Institute for Astrophysics and Space Research, Massachusetts Institute of Technology, Cambridge, MA 02139, USA
}

\date{\vspace{-1.0cm}}

\pubyear{2021}

\begin{document}
\label{firstpage}
\pagerange{\pageref{firstpage}--\pageref{lastpage}}
\maketitle

\begin{abstract}
We show that the radial velocity-variable star in the black hole candidate NGC 1850 BH1 cannot be a normal $\approx 5\,M_{\odot}$ subgiant, as was proposed, but is an overluminous stripped-envelope star with mass $\approx 1 M_{\odot}$. The result follows directly from the star's observed radius and the orbital period -- density relation for Roche lobe-filling stars: the star's density, as constrained by the observed ellipsoidal variability, is too low for its mass to exceed $\approx 1.5\,M_{\odot}$. This lower mass significantly reduces the implied mass of the unseen companion and qualitative interpretation of the system, such that a normal main-sequence companion with mass $(2.5-5)\,M_{\odot}$ is fully consistent with the data. We explore evolutionary scenarios that could produce the binary using MESA  and find that its properties can be matched by models in which a $\approx 5\,M_{\odot}$ primary loses most of its envelope to a companion and is observed in a bloated state before contracting to become a core helium burning sdOB star. This is similar to the scenario proposed to explain the binaries LB-1 and HR 6819. Though it likely does not contain a black hole, NGC 1850 BH1 provides an interesting test case for binary evolution models, particularly given its membership in a cluster of known age. 
\end{abstract}

\begin{keywords}
stars: black holes -- binaries: spectroscopic -- stars: subdwarfs
\vspace{-0.5cm}
\end{keywords}



\section{Introduction}
Characterization of the stellar-mass black hole (BH) population is a major goal of several ongoing observational surveys of the Milky Way and nearby galaxies. Despite several decades of efforts to find dormant BHs, most of the known stellar-mass BHs in the local Universe are in accreting X-ray binaries.  Only a handful of detached BH candidates remain tenable (e.g. AS 386; \citealt{Khokhlov2018}, 2MASS J05215658; \citealt{Thompson2019}, NGC 3201 \#21859 and \# 12560; \citealt{Giesers2019}, V723 Mon; \citealt{Jayasinghe2021}). None of these systems are completely unambiguous.  

Recently, \citet{Saracino2021} reported the discovery of an $11\,M_{\odot}$ BH candidate in NGC 1850, a young ($\approx $100 Myr) massive cluster in the Large Magellanic Cloud (LMC). They observed a tidally distorted star orbiting an unseen companion with period  $P_{{\rm orb}}=5.04$ days and radial velocity (RV) semi-amplitude $K = (140.4 \pm 3.3)\,{\rm km\,s^{-1}}$, corresponding to a mass function $f\left(m\right)=(1.45\pm 0.1)\,M_{\odot}$. Based primarily on the star's position in the color-magnitude diagram (CMD), they inferred that it is a subgiant with a mass $M_{{\rm donor}}\approx5\,M_{\odot}$ (we refer to the RV-variable star as the ``donor'' because it either completely or nearly fills its Roche lobe). 

Modeling the ellipsoidal variability in the binary's light curve  and assuming the donor fills its Roche lobe, \citet{Saracino2021} inferred an inclination of $i=(37.9\pm 2)$ degrees, and from this, a companion mass of $M_{2} = (11.1\pm 2)\,M_{\odot}$. A normal-star companion with such a large mass is ruled out by the object's spectrum and CMD position, so \citet{Saracino2021} conclude the unseen companion is a BH. 

No significant X-ray flux is observed from the object. Based on {\it Chandra} data, \citet{Saracino2021} find an X-ray upper limit $L_{X}\lesssim\,0.25L_{\odot}$. For context, the expected mass transfer rate for a semi-detached $5\,M_{\odot}$ subgiant is of order $10^{-7}\,M_{\odot}\,{\rm yr^{-1}}$, which exceeds the Eddington limit for an $11M_{\odot}$ BH. One would thus expect an actively accreting BH companion to have an X-ray luminosity of order $L_{{\rm edd}} \approx 1.4\times10^{39}\,{\rm erg\,s^{-1}}\approx 3.7\times10^{5}\,L_{\odot}$, about 6 orders of magnitude larger than the observed limit. Accreting BHs can, however, undergo long periods of quiescence due to disk instabilities \citep[e.g.][]{Remillard2006}. Although the lack of X-ray detection is somewhat surprising, it thus does not in itself rule out a semi-detached system hosting a BH accretor. 

The temperature and radius of the RV-variable star in NGC 1850 BH1 are quite similar to those in LB-1 and HR 6819 \citep{Liu2019, Rivinius2020}.  These objects were recently proposed to contain B stars orbiting stellar-mass BHs, but have since been convincingly argued to be mass-transfer binaries containing two luminous stars \citep[e.g.][]{Shenar2020, Bodensteiner2020, El-Badry2021}. The basic reason they were  interpreted as BH candidates is that the RV-variable B stars are not ordinary $\approx 5\,M_{\odot}$ stars near the main sequence, but undermassive (or overluminous) bloated, stripped cores with masses of only about $(0.5-1.3)\,M_{\odot}$. Interpretation of such stripped stars (which can have similar temperatures and radii to main-sequence B stars; e.g. \citealt{Irrgang2020}) as normal stars leads to large overestimates of their companions' masses, and qualitatively wrong interpretations of their evolutionary states. 

In this letter, we consider whether a similar evolutionary scenario could explain NGC 1850 BH1. We show that it can, and indeed that the alternative scenario involving a $11\,M_{\odot}$ companion is ruled out. This result is robust to modeling assumptions and relies only on the well-established relation between orbital period and density of Roche lobe-filling stars. 

\section{Basic issues}
\label{sec:issue}

The analysis of \citet{Saracino2021} relies on three main assumptions that are not necessarily justified: 
\begin{enumerate}
    \item The RV-variable star is a ``normal'' subgiant, whose evolutionary history is well described by single-star models. This leads to the inference of $M_{\rm donor}\approx 5\,M_{\odot}$. For a close binary whose components almost certainly interacted in the past, this may not be realistic.
    
    \item The donor is semi-detached, meaning that it fills its Roche lobe. Under this assumption, the authors use the observed ellipsoidal variability amplitude to estimate the binary's inclination. The relatively low inclination inferred this way ($i=38\pm 2$ deg) leads to a large implied companion mass. 
    A semi-detached donor seems reasonable given that NGC 1850 BH1 exhibits double-periodic variability, a phenomenon associated almost exclusively with mass transfer \citep{Mennickent2017}. But given the significant long-period variability, it is not clear that the inclination can be reliably constrained from the ellipsoidal variability amplitude.

    \item No luminous secondary (or disk) contributes to the observed photometry. This assumption is relevant both for interpretation of the ellipsoidal variability amplitude (a luminous secondary would dilute the true variability, making the implied inclination larger and thus the companion mass smaller), and for inference of stellar parameters from the CMD position (a luminous secondary would add light to the unresolved source).
\end{enumerate}

The most critical issue is that {\bf CMD position directly constrains the RV-variable star's temperature and radius, not its mass}. While we can roughly reproduce the radius constraint implied by the \citet{Saracino2021} analysis, we show below that the {\it mass} of the RV-variable star must be much lower than they assume -- of order $1\,M_{\odot}$, not $5\,M_{\odot}$. This strongly suggests that the star is not a normal subgiant, but a stripped product of mass transfer.

Based on comparison of the object's CMD position to isochrones, \citet{Saracino2021} infer a radius $R_{\rm donor}\approx 6 R_{\odot}$.\footnote{In particular, \citet{Saracino2021} report that their best-fit isochrones have $\log(g/\rm cm\,s^{-2})=3.57$ and $M_{\rm donor}= 5\,M_{\odot}$, implying $R_{{\rm donor}}=\left[GM_{{\rm donor}}/10^{\log(g/\rm cm\,s^{-2})}\right]^{1/2}\approx6\,R_{\odot}$.} We find consistent results when we assume that one star accounts for all the light. When we include observational uncertainties and the possibility that a second luminous star could contribute to the observed photometry, we find $4.9\leq R_{\rm donor}/R_{\odot} \leq 6.5$ (see below).

\subsection{Radius of the RV-variable star}
\label{sec:radius_constraint}
Following a similar procedure to \citet{Saracino2021}, we compared the object's HST/WFC3 photometry to a grid of MIST isochrones \citep{Choi2016}. We assumed a distance modulus of $\left(m-M\right)_{0}=18.40\pm0.05$ mag, extinction $A_V = 0.35\pm 0.1$ mag, and metallicity $\rm [Fe/H]=-0.3$  \citep{Bastian2016, Yang2018}, with a \citet{Cardelli_1989} extinction law with  $R_V =3.1$. 

We first searched for models of single stars with colors and magnitudes within 0.1 mag of the observed values, namely $\rm F336W-F438W = -0.8$ mag and $\rm F438W = 16.6$ mag. We also imposed the constraint that the effective temperature be within 2000\,K of 14,500 K, the value inferred spectroscopically by \citet{Saracino2021}. These ranges are larger than the observational uncertainties to allow for possible systematics due to e.g. an unaccounted-for companion contributing to the spectrum. This yielded a best-fit radius $R_{\rm donor} = 5.8 \pm 0.3 R_{\odot}$. The smallest donors plausibly consistent with the observed photometry and spectrum in this case have $R_{\rm donor}\approx 5.4 R_{\odot}$ with $T_{\rm eff}\approx 15,500\,\rm K$, while the largest have $R_{\rm donor}\approx 6.5 R_{\odot}$ with $T_{\rm eff}\approx 13,000\,\rm K$.

Next, we considered the possibility that an unresolved luminous companion could contribute to the photometry. In this case, the radius of the RV-variable star would be smaller, because some of the observed light would be from its companion. We considered main-sequence companions with masses $2 \leq M_2/M_{\odot} \leq 5$, between the minimum dynamically-allowed mass (Figure~\ref{fig:mass_fn}) and that above which a companion would contribute significantly to the spectrum. We still required a donor $T_{\rm eff}$ within 2000\,K of 14,500 K. In this case, we find primary and secondary combinations that can match the observed photometry down to $R_{\rm donor} = 4.9 R_{\odot}$. Because we do not know whether a luminous companion contributes to the photometry, we consider as plausible any radius between 4.9 and 6.5 $R_{\odot}$. 


\subsection{The crux of the argument: the period-density relation for semi-detached binaries}

\begin{figure}
    \centering
    \includegraphics[width=\columnwidth]{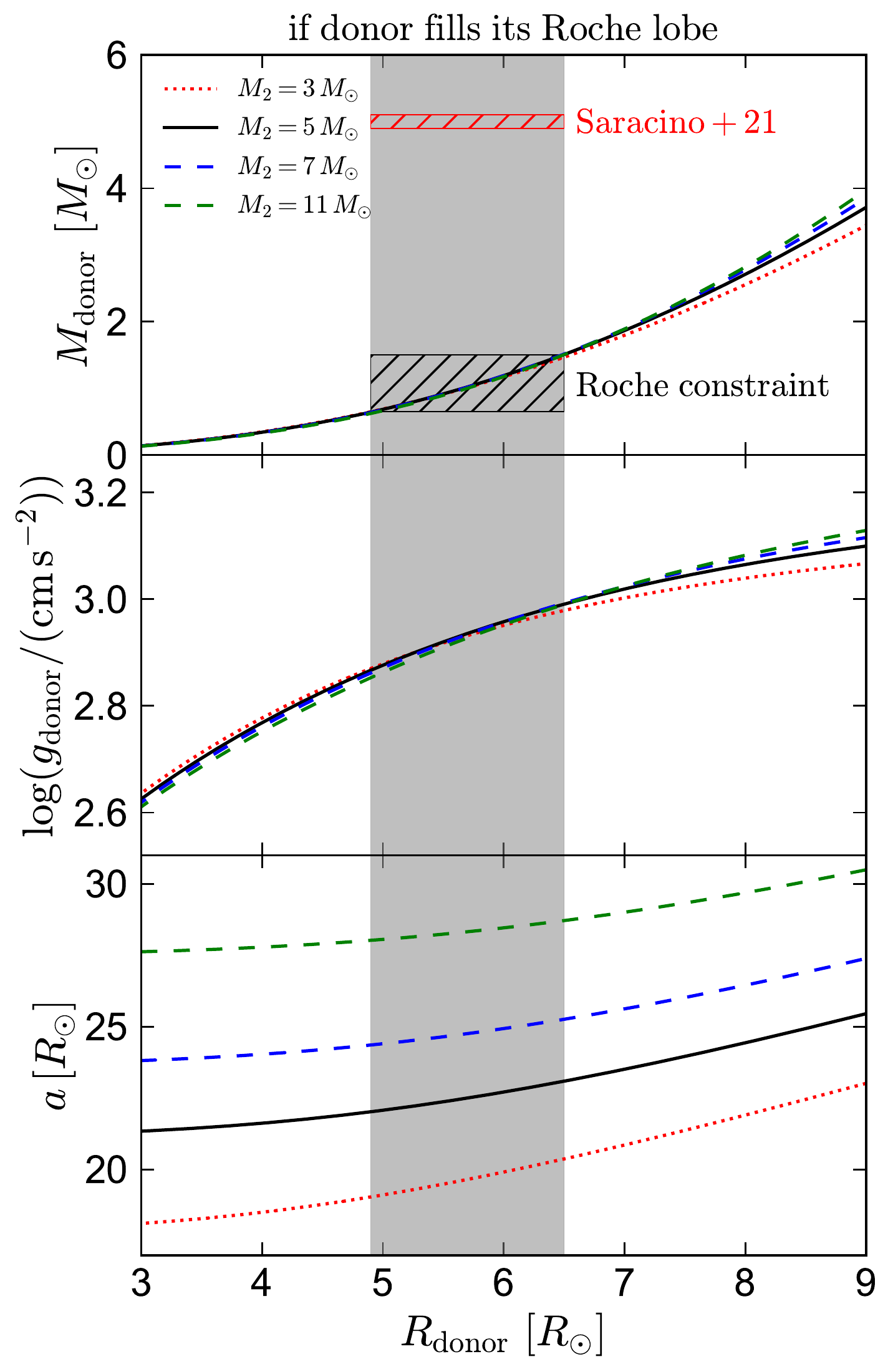}
    \caption{ Donor mass, donor surface gravity, and binary semi-major axis predicted for different donor radii if the binary is semi-detached. Shaded region shows the range of plausible donor radii, given the spectrum and the object's CMD position (Section~\ref{sec:radius_constraint}). The donor mass and surface gravity are nearly independent of the mass of the Be star for any fixed donor radius; these quantities depend only on the (known) orbital period and (reasonably well-constrained) donor radius. The donor mass is constrained to $0.65\leq M_{\rm donor}/M_{\odot} \leq 1.5$, much lower than the $5\,M_{\odot}$ assumed by \citet{Saracino2021}. This leads to a much lower companion mass (Figure~\ref{fig:mass_fn}).  }
    \label{fig:roche}
\end{figure}

\label{sec:roche_filling_eqs}
One of the most useful results in binary star physics is the fact that {\it all Roche lobe-filling stars at a given orbital period have nearly the same mean density}, irrespective of  mass or mass ratio:
\begin{align}
    \label{eq:rhostar}
    \overline{\rho}_{\rm donor} \approx 0.185\,{\rm g\,cm^{-3}}\left(P_{\rm orb}/{\rm day}\right)^{-2},
\end{align}
with very weak dependence on the mass of either component. Equation~\ref{eq:rhostar} is calculated from the \citet{Eggleton_1983} fitting formula for the Roche lobe equivalent radius and is accurate to within 6\% for mass ratios $0.01 <q < 1$, a range that comfortably encompasses all plausible mass ratios here. Here $\overline{\rho}_{\rm donor}=3M_{\rm donor} /  (4\pi R_{\rm donor}^3)$. This result has been widely used in the study of cataclysmic variables, X-ray binaries, and Algol-type binaries \citep[e.g.][]{Warner_2003}.

\begin{figure*}
    \centering
    \includegraphics[width=\textwidth]{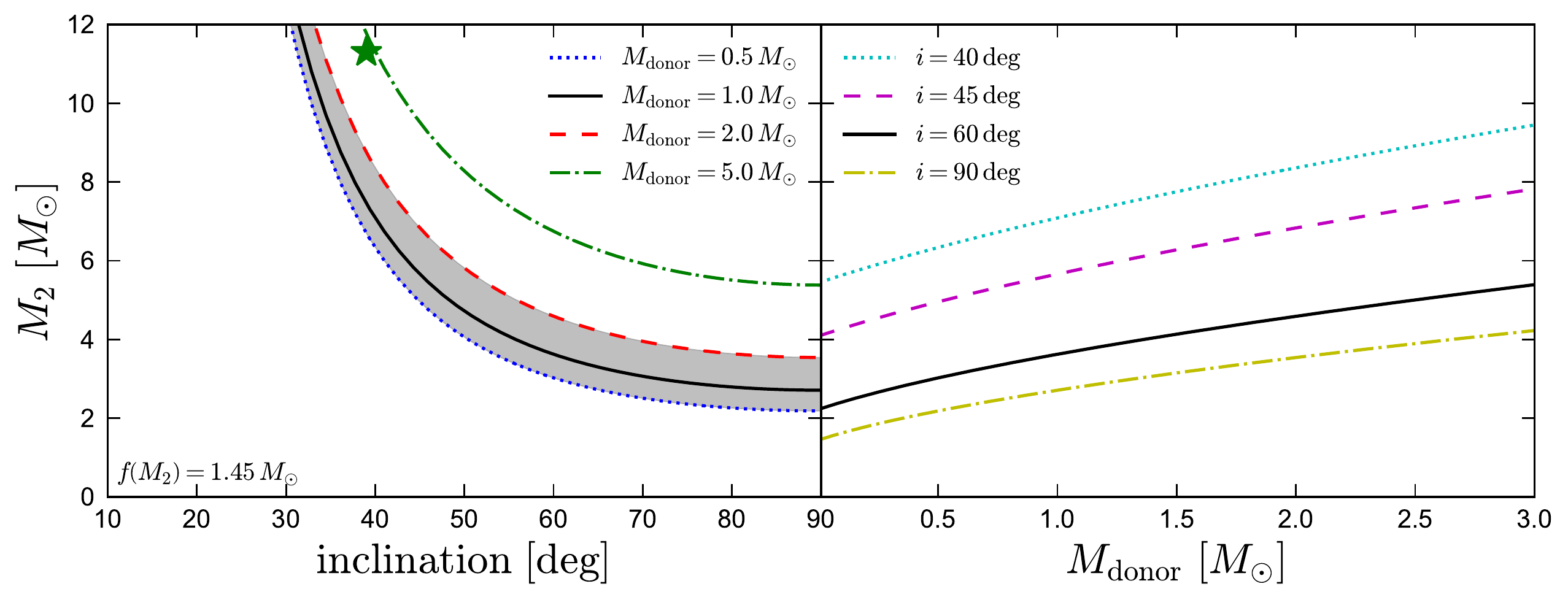}
    \caption{Implied mass of the unseen companion for different donor masses and orbital inclinations.  Green star shows the value inferred by \citet{Saracino2021}, which assumes $M_{\rm donor} = 5\,M_{\odot}$, a value ruled out by the observed radius constraint (Figure~\ref{fig:roche}). Gray shaded region gives a conservative plausible range of inclination and companion mass. The mass function is consistent with a normal-star companion with mass $2.5\lesssim M_2/M_{\odot} \lesssim 5$, which would be fainter than the donor and not easily detectable in the spectrum.}
    \label{fig:mass_fn}
\end{figure*}

From Equation~\ref{eq:rhostar}, it immediately follows that if we known the radius of a Roche-lobe filling star, we can calculate its mass. This is illustrated in Figure~\ref{fig:roche}, where we plot the allowed donor mass and surface gravity for NGC 1850 BH1 as a function of $R_{\rm donor}$. Here we use the full expression for the donor's equivalent Roche lobe radius from \citet{Eggleton_1983}, rather than the approximation in Equation~\ref{eq:rhostar}.

Irrespective of companion mass, the measured donor radius and orbital period imply that the mass of the Roche lobe-filling star is of order $1\,M_{\odot}$, ranging from $M_{\rm donor} = 0.65\,M_{\odot}$ for $R_{\rm donor}=4.9\,R_{\odot}$ to $M_{\rm donor} = 1.5\,M_{\odot}$ for  $R_{\rm donor}=6.5\,R_{\odot}$. A donor mass of $5\,M_{\odot}$ is firmly ruled out, as it would imply a radius of order $10\,R_{\odot}$ -- a donor 3 times more luminous than permitted by the observed photometry. Another way to state the problem is that a donor with $R_{\rm donor}=6\,R_{\odot}$ and  $M_{\rm donor}=5 M_{\odot}$ would only fill its Roche lobe at a period of 2.4 days, less than half the observed period.

The fact that the donor is completely or nearly Roche-lobe filling also places reasonably tight constraints on its surface gravity. Our inferred donor radii are compatible with $2.85 \lesssim \log(g/\rm cm\,s^{-2}) \lesssim 3$ if the donor fills its Roche lobe. If it does not (Section~\ref{sec:detached}), the surface gravity could be as high as $ \log(g/\rm cm\,s^{-2}) \approx 3.2$. We note that all values are lower than the $\log(g/\rm cm\,s^{-2}) =3.57$ reported by \citet{Saracino2021}. But that value was fixed based on comparison to isochrones, not measured from spectra. A robust spectroscopic measurement of $\log g$ would thus test our claim. 

One might wonder how the PHOEBE model constructed by \citet{Saracino2021} reproduces the observed ellipsoidal variability while having $M_{\rm donor}\approx 5\,M_{\odot}$. Given its orbital period and donor mass, their model must have a radius of about 10\,$R_{\odot}$, larger than permitted by the observed photometry. Such a large donor would be inconsistent with the age of NGC 1850 (given the spectroscopic temperature) and would fall about 1 mag above plausible single-star isochrones for the cluster. 

\subsection{Implications of a lower donor mass}
\label{sec:donor_mass}
With a lower-mass donor, the companion masses implied by the observed RVs and mass function decrease by about a factor of 2 at fixed inclination. This is illustrated in Figure~\ref{fig:mass_fn}. 

A main-sequence companion with, say, $M_2=4\,M_{\odot}$ and inclination $i\approx 60$ deg is consistent with the RVs. For such a companion, the lack of observed eclipses limits the inclination to $i \lesssim 67\,\rm deg$, with weak dependence on the donor and companion radii. We do not think it is obvious that the system is semi-detached and thus do not attempt to place a tight limit on the inclination from the amplitude of ellipsoidal variability. However, we note that the implied inclination is generally more edge-on with a luminous companion than with a dark one, because the ellipsoidal variability is diluted by its light. 

A main-sequence secondary would have temperature within a few thousand K of the donor and a somewhat smaller radius, with luminosity ranging from 10 to 100\% of the donor's luminosity. Concretely, a binary with $M_{\rm donor}\approx 1.1\,M_{\odot}$, $R_{\rm donor}\approx 5.7\,R_{\odot}$, $M_2\approx 3.5\,M_{\odot}$, and $i\approx 60$ deg is one configuration that would match the observed RVs and light curve. It may seem strange that a normal-star companion could escape detection in the spectrum. The key point here is that accretion of the donor's envelope can spin up the accretor, smearing out its spectral lines. It is for this reason that the luminous companions to LB-1 and HR 6819 were difficult to detect even with very high-resolution, high-SNR spectra. We note that if a luminous secondary does contribute significantly to the observed spectra, this could lead to biases in the inferred donor temperature and RVs.

\subsection{What if the system is detached?}
\label{sec:detached}
We suspect that the donor fills its Roche lobe, given the observed double-periodic variability. If it does not, its mass would be larger than the Roche-filling limit shown in Figure~\ref{fig:roche}. Given an observed peak-to-peak ellipsoidal variability amplitude of about 12\% in the $V$ band (after removing contamination from an unrelated nearby star; see \citealt{Saracino2021}), and assuming a typical inclination of 60 deg, we infer $R_{\rm donor}/R_{\rm Roche\,lobe}\gtrsim 0.85$ (see \citealt{El-Badry2021_cvs}, their Figure 10). A Roche lobe filling factor of 0.85 would inflate the implied masses in Figure~\ref{fig:roche} by a factor of 1.63; leading to $1.06 \leq M_{\rm donor}/M_{\odot} \leq 2.45$ instead of $0.65 \leq M_{\rm donor}/M_{\odot} \leq 1.5$. Such a scenario never leads to a significantly larger implied $M_2$ than when one assumes the system is Roche-lobe filling, because the increase in $M_2$ due to the larger $M_{\rm donor}$ is offset by a decrease due to the higher implied inclination. Thus, any $M_2\gtrsim 6\,M_{\odot}$ is ruled out.

\section{Evolutionary history}
\label{sec:history}

To investigate the binary's possible evolutionary history, we searched in a suite of binary evolution calculations (El-Badry et al., in prep) for models that at some point in their evolution match the observed orbital period and donor effective temperature and surface gravity.  The models were calculated with MESA \citep[Modules for Experiments in Stellar Astrophysics, version \texttt{r15140};][]{Paxton_2011, Paxton_2013, Paxton_2015, Paxton_2018, Paxton_2019}, which simultaneously solves the 1D stellar structure equations for both stars, while accounting for mass and angular momentum transfer using simplified prescriptions.

The \texttt{MESAbinary} module is described by \citet{Paxton_2015}. We used the \texttt{evolve\_both\_stars} inlists in the MESA test suite as a starting point for our calculations, and most inlist parameters are set to their default values.  Roche lobe radii are computed using the fit of \citet{Eggleton_1983}. Mass transfer rates in Roche lobe overflowing systems are determined following the prescription of \citet{Kolb_1990}, as described in Equations 13-18 of \citet{Paxton_2015}. The orbital separation evolves such that the total angular momentum is conserved when mass is lost or transferred to a companion, as described by \citet{Paxton_2015}. We set \texttt{mass\_transfer\_beta = 0.4}, \texttt{mass\_transfer\_delta = 0.25}, \texttt{mass\_transfer\_gamma = $\sqrt{2}$}. That is, we assume that when the primary loses mass, 40\% escapes the system from the vicinity of the accretor as a fast wind,  25\% escapes through a circumbinary ring with radius $2a$, and the rest is accreted by the secondary.\footnote{In reality, these parameters are expected to be functions of the mass-transfer rate and other instantaneous properties of the binary. These choices should thus be regarded as an effective model of a more complicated mass transfer process; likely, one of several possible combinations that can produce the right amount of angular momentum loss to match the observed period. However, some form of non-conservative angular momentum loss is needed to match the observed period.} We set the metallicity to $Z=0.008$.

\begin{figure*}
    \centering
    \includegraphics[width=\textwidth]{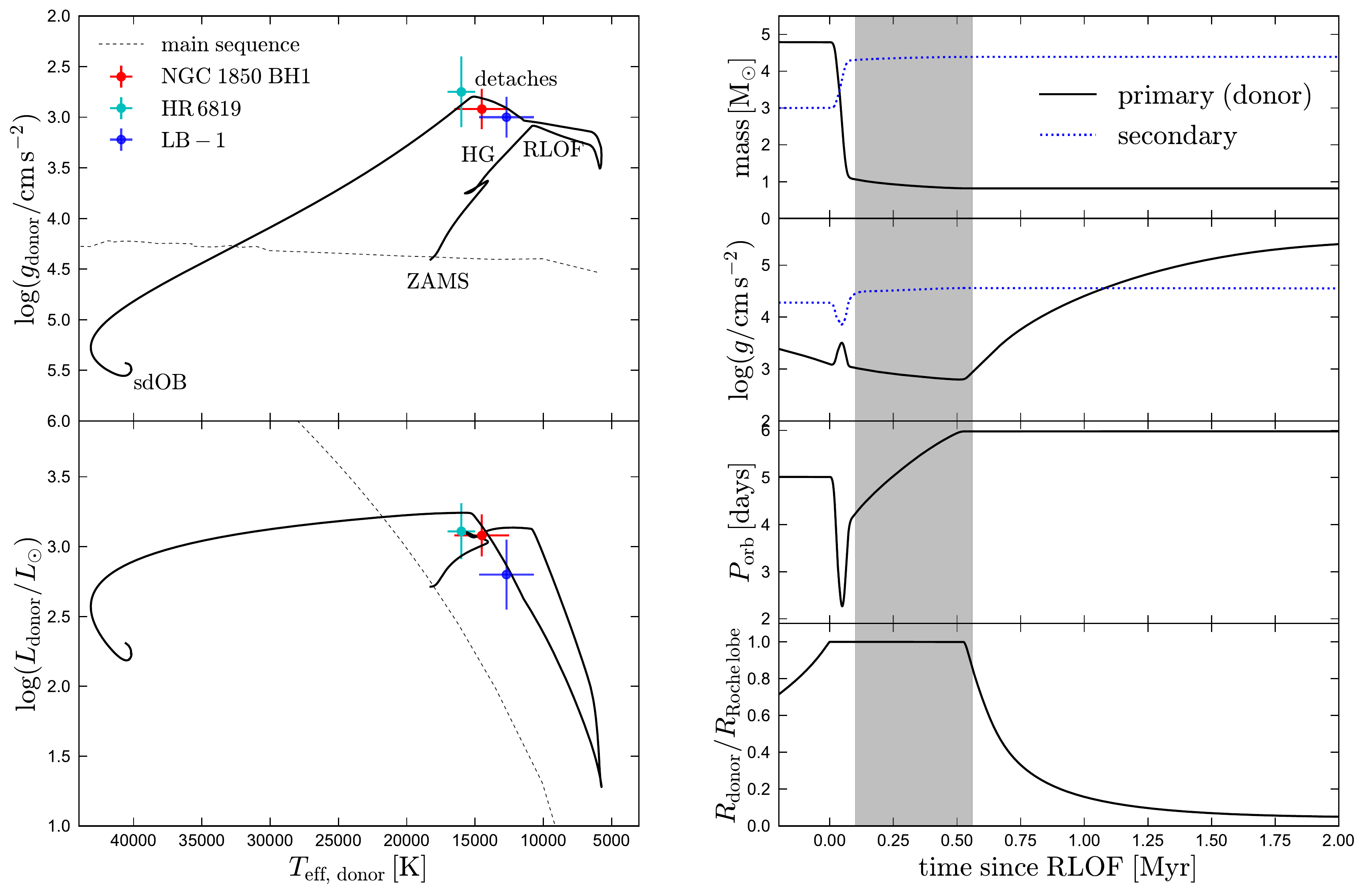}
    \caption{MESA calculation for a binary with initial masses of 4.8 and 3 $M_{\odot}$ and initial period of 5 days. Left panels show the evolution of the primary, which begins at the zero-age main sequence (ZAMS). Roche lobe overflow (RLOF) occurs in the Hertzsprung gap (HG) at $t =102\,\rm Myr$ and is followed by a brief period of thermal-timescale mass transfer, during which the donor cools and loses most of its envelope. The stripped donor then begins to heat up, reaching $T_{\rm eff}\approx 14,500\,\rm K$ and $\log(g/\rm cm\,s^{-2}) \approx 3$ at $P_{\rm orb}= 4-6$ days. Mass transfer ceases after 0.55 Myr, when the donor has lost most of its envelope. The donor then contracts and heats up further, settling as a core helium burning sdOB star. Shading (right panels) marks the period during which the binary resembles NGC 1850 BH1. In the left panels, we also show HR 6819 and LB-1. These are likely in a qualitatively similar evolutionary state to NGC 1850 BH1 but are detached and have wider orbits, so they are not matched by exactly the same MESA model. }
    \label{fig:mesa}
\end{figure*}

One satisfactory model is shown in Figure~\ref{fig:mesa}. The model begins with an initial primary mass of $4.8\,M_{\odot}$  and a companion mass $3\,M_{\odot}$ in a 5 day orbit. Because the binary in a 100-Myr old cluster, the primary's initial mass must have been about $5\,M_{\odot}$, such that it just recently terminated its main-sequence evolution. Mass transfer begins after 102 Myr, when the primary is a subgiant, and initially occurs on a thermal timescale. During this period, the mass transfer rate is large (between $10^{-5}$ and $10^{-4}\,M_{\odot}\,\rm yr^{-1}$), so it is unlikely that the secondary would be able to retain all the transferred mass. Once most of the primary's envelope has been stripped, it begins to heat up, passing the observed $T_{\rm eff}$ and the $\log g$ inferred from Figure~\ref{fig:roche} while still filling its Roche lobe. After a 0.5 Myr period during which it resembles NGC 1850 BH1, the donor begins to contract and heat more rapidly, finally settling as a core helium burning sdOB star.

We stress that the model in Figure~\ref{fig:mesa} represents just one illustrative evolutionary scenario for producing a stripped star with a normal-star companion. Additional models were recently explored by \citet{Stevance2021}. The nature of the secondary (its initial mass, the efficiency with which it retained mass lost by the primary, how much it was spun-up, etc.) are as yet quite uncertain. Observational measurement of these quantities would help constrain the model further.  

\section{Discussion}
\label{sec:discussion}
We have shown that the RV-variable star in NGC 1850 BH1 cannot be a normal $5\,M_{\odot}$ subgiant, but must instead be a $\approx 1\,M_{\odot}$ stripped star. A $5\,M_{\odot}$ star would simply be too compact to produce the observed ellipsoidal distortion at this period, irrespective of the companion mass (Figure~\ref{fig:roche}). This then also implies a lower companion mass, making the data consistent with a luminous companion (Figure~\ref{fig:mass_fn}). Such a system can be produced by mass transfer in an Algol-type binary (Figure~\ref{fig:mesa}). In this scenario, accretion would likely spin up the companion, smearing out its absorption lines and making its contribution to the observed spectrum  difficult to detect.

The data rule out any companion more massive than $\approx 6\,M_{\odot}$. Because we have not yet detected another luminous star in the system, we have not ruled out the possibility of a lower-mass (say, $5\,M_{\odot}$) BH companion. Our results do show, however, that the binary's properties can be explained without invoking a BH. Given the rarity of BHs in this type of system -- which would be just about to become a LMXB or IMXB accreting near the Eddington limit -- and the lack of observed X-rays, the more banal explanation seems  more probable. 

NGC 1850 BH1 is quite similar to LB-1 and HR 6819, two previously proposed BH candidates, with two main differences.  First, the companions in LB-1 and HR 6819 are Be stars with emission lines, and no emission lines are observed in this system. Second, this system is significantly more compact, with an orbital period of only 5 days, compared to 40 days in HR 6819 and 79 days in LB-1. The emission lines in HR 6819 and LB-1 are proposed to originate from a decretion disk around the secondary, potentially a result of spin-up by mass transfer. Lack of a disk in NGC 1850 BH1 might imply that the accretor was not spun-up sufficiently to reach critical rotation. This could occur if most of the donor's mass was lost on a timescale too short for the secondary to retain it, which would also help explain the shorter orbital period. We also note that Be star disks are known to grow and fade \citep[e.g.][]{Rivinius2013}, and a disk in this compact system would necessarily be much smaller than in typical Be stars, so it is also possible that a faint disk exists but was not detected. 

There is one more important wrinkle: NGC 1850 BH1 is a double-periodic variable (DPV; e.g. \citealt{Mennickent2003}): in addition to ellipsoidal variability with $P_{\rm orb} = 5.04$ days, the object's OGLE light curve reveals stable, roughly sinusoidal variability with period 156 days \citep{Poleski2010}.  The $I$-band peak-to-peak amplitudes of the long- and short-period modulations are similar. The long-period variability in DPVs is very likely related to the mass transfer process, but its precise origin remains unclear. Despite having high mass-transfer rates, many DPVs do not show obvious emission lines, though there is often subtle infilling of the Balmer lines \citep{Mennickent2005}. DPVs that have been subject to spectroscopic follow-up all contain rapidly-rotating B type accretors near the main-sequence and evolved, stripped donors: they are ``hot-Algols'' \citep{Mennickent2017}, precisely the scenario we propose for NGC 1850 BH1. 

NGC 1850 BH1 will be particularly informative to study and model in more detail because unlike other systems in its class, it is in a  cluster of known age. This means that the initial mass of the donor is well-constrained to $\approx 5 M_{\odot}$. We are optimistic that further study of the system, including phase-resolved spectroscopy and measurement of detailed abundances and rotation rates, will shine further light on its evolutionary history and that of other similar objects.

\section*{Acknowledgements}
We thank the referee, Christopher Tout, for a constructive report, and Sara Saracino, Sebastian Kamann, Nate Bastian, Christopher Usher, Ivan Cabrera, Mark Gieles, Tom Prince, Jim Fuller, Charlie Conroy, and Rohan Naidu for useful comments.

\section*{Data Availability}
All data used in this study are publicly available.



\bibliographystyle{mnras}

\begin{thebibliography}{}
\makeatletter
\relax
\def\mn@urlcharsother{\let\do\@makeother \do\$\do\&\do\#\do\^\do\_\do\%\do\~}
\def\mn@doi{\begingroup\mn@urlcharsother \@ifnextchar [ {\mn@doi@}
  {\mn@doi@[]}}
\def\mn@doi@[#1]#2{\def\@tempa{#1}\ifx\@tempa\@empty \href
  {http://dx.doi.org/#2} {doi:#2}\else \href {http://dx.doi.org/#2} {#1}\fi
  \endgroup}
\def\mn@eprint#1#2{\mn@eprint@#1:#2::\@nil}
\def\mn@eprint@arXiv#1{\href {http://arxiv.org/abs/#1} {{\tt arXiv:#1}}}
\def\mn@eprint@dblp#1{\href {http://dblp.uni-trier.de/rec/bibtex/#1.xml}
  {dblp:#1}}
\def\mn@eprint@#1:#2:#3:#4\@nil{\def\@tempa {#1}\def\@tempb {#2}\def\@tempc
  {#3}\ifx \@tempc \@empty \let \@tempc \@tempb \let \@tempb \@tempa \fi \ifx
  \@tempb \@empty \def\@tempb {arXiv}\fi \@ifundefined
  {mn@eprint@\@tempb}{\@tempb:\@tempc}{\expandafter \expandafter \csname
  mn@eprint@\@tempb\endcsname \expandafter{\@tempc}}}

\bibitem[\protect\citeauthoryear{{Bastian} et~al.,}{{Bastian}
  et~al.}{2016}]{Bastian2016}
{Bastian} N.,  et~al., 2016, \mn@doi [\mnras] {10.1093/mnrasl/slw067}, \href
  {https://ui.adsabs.harvard.edu/abs/2016MNRAS.460L..20B} {460, L20}

\bibitem[\protect\citeauthoryear{{Bodensteiner} et~al.,}{{Bodensteiner}
  et~al.}{2020}]{Bodensteiner2020}
{Bodensteiner} J.,  et~al., 2020, \mn@doi [\aap] {10.1051/0004-6361/202038682},
  \href {https://ui.adsabs.harvard.edu/abs/2020A&A...641A..43B} {641, A43}

\bibitem[\protect\citeauthoryear{{Cardelli}, {Clayton}  \& {Mathis}}{{Cardelli}
  et~al.}{1989}]{Cardelli_1989}
{Cardelli} J.~A.,  {Clayton} G.~C.,   {Mathis} J.~S.,  1989, \mn@doi [\apj]
  {10.1086/167900}, \href
  {https://ui.adsabs.harvard.edu/abs/1989ApJ...345..245C} {345, 245}

\bibitem[\protect\citeauthoryear{{Choi}, {Dotter}, {Conroy}, {Cantiello},
  {Paxton}  \& {Johnson}}{{Choi} et~al.}{2016}]{Choi2016}
{Choi} J.,  {Dotter} A.,  {Conroy} C.,  {Cantiello} M.,  {Paxton} B.,
  {Johnson} B.~D.,  2016, \mn@doi [\apj] {10.3847/0004-637X/823/2/102}, \href
  {https://ui.adsabs.harvard.edu/abs/2016ApJ...823..102C} {823, 102}

\bibitem[\protect\citeauthoryear{{Eggleton}}{{Eggleton}}{1983}]{Eggleton_1983}
{Eggleton} P.~P.,  1983, \mn@doi [\apj] {10.1086/160960}, \href
  {https://ui.adsabs.harvard.edu/abs/1983ApJ...268..368E} {268, 368}

\bibitem[\protect\citeauthoryear{{El-Badry} \& {Quataert}}{{El-Badry} \&
  {Quataert}}{2021}]{El-Badry2021}
{El-Badry} K.,  {Quataert} E.,  2021, \mn@doi [\mnras] {10.1093/mnras/stab285},
  \href {https://ui.adsabs.harvard.edu/abs/2021MNRAS.502.3436E} {502, 3436}

\bibitem[\protect\citeauthoryear{{El-Badry}, {Rix}, {Quataert}, {Kupfer}  \&
  {Shen}}{{El-Badry} et~al.}{2021}]{El-Badry2021_cvs}
{El-Badry} K.,  {Rix} H.-W.,  {Quataert} E.,  {Kupfer} T.,   {Shen} K.~J.,
  2021, \mn@doi [\mnras] {10.1093/mnras/stab2583}, \href
  {https://ui.adsabs.harvard.edu/abs/2021MNRAS.508.4106E} {508, 4106}

\bibitem[\protect\citeauthoryear{{Giesers} et~al.,}{{Giesers}
  et~al.}{2019}]{Giesers2019}
{Giesers} B.,  et~al., 2019, \mn@doi [\aap] {10.1051/0004-6361/201936203},
  \href {https://ui.adsabs.harvard.edu/abs/2019A&A...632A...3G} {632, A3}

\bibitem[\protect\citeauthoryear{{Irrgang}, {Geier}, {Kreuzer}, {Pelisoli}  \&
  {Heber}}{{Irrgang} et~al.}{2020}]{Irrgang2020}
{Irrgang} A.,  {Geier} S.,  {Kreuzer} S.,  {Pelisoli} I.,   {Heber} U.,  2020,
  \mn@doi [\aap] {10.1051/0004-6361/201937343}, \href
  {https://ui.adsabs.harvard.edu/abs/2020A&A...633L...5I} {633, L5}

\bibitem[\protect\citeauthoryear{{Jayasinghe} et~al.,}{{Jayasinghe}
  et~al.}{2021}]{Jayasinghe2021}
{Jayasinghe} T.,  et~al., 2021, \mn@doi [\mnras] {10.1093/mnras/stab907}, \href
  {https://ui.adsabs.harvard.edu/abs/2021MNRAS.504.2577J} {504, 2577}

\bibitem[\protect\citeauthoryear{{Khokhlov} et~al.,}{{Khokhlov}
  et~al.}{2018}]{Khokhlov2018}
{Khokhlov} S.~A.,  et~al., 2018, \mn@doi [\apj] {10.3847/1538-4357/aab49d},
  \href {https://ui.adsabs.harvard.edu/abs/2018ApJ...856..158K} {856, 158}

\bibitem[\protect\citeauthoryear{{Kolb} \& {Ritter}}{{Kolb} \&
  {Ritter}}{1990}]{Kolb_1990}
{Kolb} U.,  {Ritter} H.,  1990, \aap, \href
  {https://ui.adsabs.harvard.edu/abs/1990A&A...236..385K} {236, 385}

\bibitem[\protect\citeauthoryear{{Liu} et~al.,}{{Liu} et~al.}{2019}]{Liu2019}
{Liu} J.,  et~al., 2019, \mn@doi [\nat] {10.1038/s41586-019-1766-2}, \href
  {https://ui.adsabs.harvard.edu/abs/2019Natur.575..618L} {575, 618}

\bibitem[\protect\citeauthoryear{{Mennickent}}{{Mennickent}}{2017}]{Mennickent2017}
{Mennickent} R.~E.,  2017, \mn@doi [Serbian Astronomical Journal]
  {10.2298/SAJ1794001M}, \href
  {https://ui.adsabs.harvard.edu/abs/2017SerAJ.194....1M} {194, 1}

\bibitem[\protect\citeauthoryear{{Mennickent}, {Pietrzy{\'n}ski}, {Diaz}  \&
  {Gieren}}{{Mennickent} et~al.}{2003}]{Mennickent2003}
{Mennickent} R.~E.,  {Pietrzy{\'n}ski} G.,  {Diaz} M.,   {Gieren} W.,  2003,
  \mn@doi [\aap] {10.1051/0004-6361:20030106}, \href
  {https://ui.adsabs.harvard.edu/abs/2003A&A...399L..47M} {399, L47}

\bibitem[\protect\citeauthoryear{{Mennickent}, {Cidale}, {D{\'\i}az},
  {Pietrzy{\'n}ski}, {Gieren}  \& {Sabogal}}{{Mennickent}
  et~al.}{2005}]{Mennickent2005}
{Mennickent} R.~E.,  {Cidale} L.,  {D{\'\i}az} M.,  {Pietrzy{\'n}ski} G.,
  {Gieren} W.,   {Sabogal} B.,  2005, \mn@doi [\mnras]
  {10.1111/j.1365-2966.2005.08718.x}, \href
  {https://ui.adsabs.harvard.edu/abs/2005MNRAS.357.1219M} {357, 1219}

\bibitem[\protect\citeauthoryear{{Paxton}, {Bildsten}, {Dotter}, {Herwig},
  {Lesaffre}  \& {Timmes}}{{Paxton} et~al.}{2011}]{Paxton_2011}
{Paxton} B.,  {Bildsten} L.,  {Dotter} A.,  {Herwig} F.,  {Lesaffre} P.,
  {Timmes} F.,  2011, \mn@doi [\apjs] {10.1088/0067-0049/192/1/3}, \href
  {https://ui.adsabs.harvard.edu/abs/2011ApJS..192....3P} {192, 3}

\bibitem[\protect\citeauthoryear{{Paxton} et~al.,}{{Paxton}
  et~al.}{2013}]{Paxton_2013}
{Paxton} B.,  et~al., 2013, \mn@doi [\apjs] {10.1088/0067-0049/208/1/4}, \href
  {https://ui.adsabs.harvard.edu/abs/2013ApJS..208....4P} {208, 4}

\bibitem[\protect\citeauthoryear{{Paxton} et~al.,}{{Paxton}
  et~al.}{2015}]{Paxton_2015}
{Paxton} B.,  et~al., 2015, \mn@doi [\apjs] {10.1088/0067-0049/220/1/15}, \href
  {https://ui.adsabs.harvard.edu/abs/2015ApJS..220...15P} {220, 15}

\bibitem[\protect\citeauthoryear{{Paxton} et~al.,}{{Paxton}
  et~al.}{2018}]{Paxton_2018}
{Paxton} B.,  et~al., 2018, \mn@doi [\apjs] {10.3847/1538-4365/aaa5a8}, \href
  {https://ui.adsabs.harvard.edu/abs/2018ApJS..234...34P} {234, 34}

\bibitem[\protect\citeauthoryear{{Paxton} et~al.,}{{Paxton}
  et~al.}{2019}]{Paxton_2019}
{Paxton} B.,  et~al., 2019, \mn@doi [\apjs] {10.3847/1538-4365/ab2241}, \href
  {https://ui.adsabs.harvard.edu/abs/2019ApJS..243...10P} {243, 10}

\bibitem[\protect\citeauthoryear{{Poleski}, {Soszy{\'n}ski}, {Udalski},
  {Szyma{\'n}ski}, {Kubiak}, {Pietrzy{\'n}ski}, {Wyrzykowski}  \&
  {Ulaczyk}}{{Poleski} et~al.}{2010}]{Poleski2010}
{Poleski} R.,  {Soszy{\'n}ski} I.,  {Udalski} A.,  {Szyma{\'n}ski} M.~K.,
  {Kubiak} M.,  {Pietrzy{\'n}ski} G.,  {Wyrzykowski} {\L}.,   {Ulaczyk} K.,
  2010, \actaa, \href {https://ui.adsabs.harvard.edu/abs/2010AcA....60..179P}
  {60, 179}

\bibitem[\protect\citeauthoryear{{Remillard} \& {McClintock}}{{Remillard} \&
  {McClintock}}{2006}]{Remillard2006}
{Remillard} R.~A.,  {McClintock} J.~E.,  2006, \mn@doi [\araa]
  {10.1146/annurev.astro.44.051905.092532}, \href
  {https://ui.adsabs.harvard.edu/abs/2006ARA&A..44...49R} {44, 49}

\bibitem[\protect\citeauthoryear{{Rivinius}, {Carciofi}  \&
  {Martayan}}{{Rivinius} et~al.}{2013}]{Rivinius2013}
{Rivinius} T.,  {Carciofi} A.~C.,   {Martayan} C.,  2013, \mn@doi [\aapr]
  {10.1007/s00159-013-0069-0}, \href
  {https://ui.adsabs.harvard.edu/abs/2013A&ARv..21...69R} {21, 69}

\bibitem[\protect\citeauthoryear{{Rivinius}, {Baade}, {Hadrava}, {Heida}  \&
  {Klement}}{{Rivinius} et~al.}{2020}]{Rivinius2020}
{Rivinius} T.,  {Baade} D.,  {Hadrava} P.,  {Heida} M.,   {Klement} R.,  2020,
  \mn@doi [\aap] {10.1051/0004-6361/202038020}, \href
  {https://ui.adsabs.harvard.edu/abs/2020A&A...637L...3R} {637, L3}

\bibitem[\protect\citeauthoryear{{Saracino} et~al.,}{{Saracino}
  et~al.}{2021}]{Saracino2021}
{Saracino} S.,  et~al., 2021, \mn@doi [\mnras] {10.1093/mnras/stab3159}, \href
  {https://ui.adsabs.harvard.edu/abs/2021MNRAS.tmp.2924S} {}

\bibitem[\protect\citeauthoryear{{Shenar} et~al.,}{{Shenar}
  et~al.}{2020}]{Shenar2020}
{Shenar} T.,  et~al., 2020, \mn@doi [\aap] {10.1051/0004-6361/202038275}, \href
  {https://ui.adsabs.harvard.edu/abs/2020A&A...639L...6S} {639, L6}

\bibitem[\protect\citeauthoryear{{Stevance}, {Parsons}  \&
  {Eldridge}}{{Stevance} et~al.}{2021}]{Stevance2021}
{Stevance} H.~F.,  {Parsons} S.~G.,   {Eldridge} J.~J.,  2021, arXiv e-prints,
  \href {https://ui.adsabs.harvard.edu/abs/2021arXiv211200015S} {p.
  arXiv:2112.00015}

\bibitem[\protect\citeauthoryear{{Thompson} et~al.,}{{Thompson}
  et~al.}{2019}]{Thompson2019}
{Thompson} T.~A.,  et~al., 2019, \mn@doi [Science] {10.1126/science.aau4005},
  \href {https://ui.adsabs.harvard.edu/abs/2019Sci...366..637T} {366, 637}

\bibitem[\protect\citeauthoryear{{Warner}}{{Warner}}{2003}]{Warner_2003}
{Warner} B.,  2003, {Cataclysmic Variable Stars},
  \mn@doi{10.1017/CBO9780511586491.
}

\bibitem[\protect\citeauthoryear{{Yang}, {Li}, {Deng}, {de Grijs}  \&
  {Milone}}{{Yang} et~al.}{2018}]{Yang2018}
{Yang} Y.,  {Li} C.,  {Deng} L.,  {de Grijs} R.,   {Milone} A.~P.,  2018,
  \mn@doi [\apj] {10.3847/1538-4357/aabe26}, \href
  {https://ui.adsabs.harvard.edu/abs/2018ApJ...859...98Y} {859, 98}

\makeatother
\end{thebibliography}



%
%


\bsp	
\label{lastpage}
\end{document}